\documentclass{aa}
\usepackage{graphicx,natbib}

\usepackage{txfonts}
\begin{document}
\title{GRB\,050904 at redshift 6.3: observations of the oldest cosmic explosion after the Big Bang%
\thanks{Based on observations carried out at ESO telescopes under
program Id 075.D-0787. We also used data from Telescopio Nazionale Galileo and 
Centro Astron\'omico Hispano  Alem\'an.}}

\author{G. Tagliaferri\inst{1}
\and L.A.~Antonelli\inst{2} \and G. Chincarini\inst{1,3}
\and A.~Fern\'andez-Soto\inst{4} \and D.~Malesani\inst{5}
\and M.~Della Valle\inst{6} \and P.~D'Avanzo\inst{1,7}
\and A.~Grazian\inst{2} \and V.~Testa\inst{2}
\and S.~Campana\inst{1} \and S.~Covino\inst{1}
\and F.~Fiore\inst{2} \and L.~Stella\inst{2}
\and A.J.~Castro-Tirado\inst{8} \and J.~Gorosabel\inst{8}
\and D.N.~Burrows\inst{9} \and M.~Capalbi\inst{10} \and G.~Cusumano\inst{11}
\and M.L.~Conciatore\inst{2} \and V.~D'Elia\inst{2}
\and P.~Filliatre\inst{12,13} \and D.~Fugazza\inst{1} \and N.~Gehrels\inst{14}
\and P.~Goldoni\inst{12,13}
\and D.~Guetta\inst{2} \and S. Guziy\inst{8}
\and E.V.~Held\inst{15} \and K. Hurley\inst{16} \and G.L.~Israel\inst{2}
\and M.~Jel\'{\i}nek\inst{8} \and D.~Lazzati\inst{17} \and A.~L\'opez-Echarri\inst{18}
\and A.~Melandri\inst{2,19}
\and I.F.~Mirabel\inst{20} \and M.~Moles\inst{8} \and A.~Moretti\inst{1} \and K.O.~Mason\inst{21}
\and J.~Nousek\inst{9} \and J.~Osborne\inst{22}
\and L.J.~Pellizza\inst{23} \and R.~Perna\inst{17}
\and S.~Piranomonte\inst{2} \and L.~Piro\inst{24}
\and A.~de~Ugarte Postigo\inst{8} \and P.~Romano\inst{1}
}

\offprints{G. Tagliaferri,\\\email{tagliaferri@merate.mi.astro.it}}

\institute{ 
INAF-Osservatorio Astronomico di Brera, via E. Bianchi 46, I-23807 Merate (Lc), Italy
\and        
INAF-Osservatorio Astronomico di Roma, via di Frascati 33, I-00040 Monteporzio Catone (Roma), Italy
\and        
Universit\`a degli studi di Milano-Bicocca, Dipartimento di Fisica, piazza delle Scienze 3, I-20126 Milano, Italy
\and        
Observatori Astronomic, Universitat de Valencia, Aptdo. Correos 22085, Valencia, 46071, Spain
\and        
International School for Advanced Studies (SISSA-ISAS), via Beirut 2-4, I-34014 Trieste, Italy
\and        
INAF-Osservatorio Astrofisico di Arcetri, largo E. Fermi 5, I-50125 Firenze, Italy%
\and        
Dipartimento di Fisica e Matematica, Universit\`a dell'Insubria, via Valleggio 11, I-22100 Como, Italy
\and        
Inst{\'\i}tuto de Astrof{\'\i}sica de Andaluc{\'\i}a (IAA-CSIC), P.O. Box 03004, E-18080 Granada, Spain
\and       
Department of Astronomy \& Astrophysics, Pennsylvania State University, State College, PA 16802, USA
\and        
ASI Science Data Center, via G. Galilei 5, I-00044 Frascati (Roma), Italy
\and       
INAF-IASF, sezione di Palermo, via U. La Malfa 153, I-90146 Palermo, Italy
\and      
Laboratoire Astroparticule et Cosmologie, UMR 7164, 11 Place Marcelin Berthelot, F-75231 Paris Cedex 05, France 
\and      
Service d'Astrophysique, DSM/DAPNIA, CEA Saclay, F-91911 Gif-sur-Yvette Cedex, France
\and      
NASA, Goddard Space Flight Center, Greenbelt, MD 20771, USA
\and      
INAF-Osservatorio Astronomico di Padova, vicolo dell'Osservatorio 5, I-35122 Padova, Italy
\and      
University of California, Berkeley, Space Sciences Laboratory, Berkeley, CA 94720-7450, USA
\and       
JILA, University of Colorado, 440 UCB, Boulder CO 80309-0440, USA
\and       
Inst{\'\i}tuto de Astrof{\'\i}sica de Canarias, C/ V{\'\i}a L\'actea s/n, E-38200 La Laguna (Tenerife), Spain
\and      
Universit\`a degli Studi di Cagliari, Dipartimento di Fisica, I-09042 Monserrato (Ca), Italy
\and      
European Southern Observatory - Vitacura, Casilla 19001, Santiago 19, Chile 
\and      
MSSL, University College London, Holmbury St. Mary, Dorking, RH5 6NT Surrey, UK
\and      
X-Ray \& Observational Astronomy Group, Dept. of Physics \& Astronomy, University of Leicester, Leicester LE1 7RH, UK
\and      
AIM (UMR 7158 CEA/CNRS/Universit\'e Paris 7), Service d'Astrophysique, CEA Saclay, F-91191 Gif-sur-Yvette, France
\and      
INAF-IASF, sezione di Roma, via Fosso del Cavaliere 100, I-00133 Roma, Italy
}
\date{}

\abstract{

We present optical and near-infrared observations of the afterglow of
the gamma-ray burst GRB\,050904. We derive a photometric redshift $z =
6.3$, estimated from the presence of the Lyman break falling between the
$I$ and $J$ filters.  This is by far the most distant GRB known to date.
Its isotropic-equivalent energy is $3.4 \times 10^{53}$~erg in the
rest-frame 110-1100~keV energy band.  Despite the high redshift, both
the prompt and the afterglow emission are not peculiar with respect to
other GRBs.  We find a break in the $J$-band light curve at $t_{\rm b} =
2.6 \pm 1.0$~d (observer frame). If we assume this is the jet break, we
derive a beaming-corrected energy $E_\gamma \sim (4 \div 12) \times 10^{51}$~erg.
This limit shows that GRB\,050904 is consistent with the Amati and
Ghirlanda relations.  This detection is consistent with the expected
number of GRBs at $z > 6$ and shows that GRBs are a powerful tool to
study the star formation history up to very high redshift.
}

\authorrunning{Tagliaferri et al.}
\titlerunning{GRB\,050904 at redshift 6.3}

\maketitle


\section{Introduction}
\label{sec:intro}

Gamma-ray bursts (GRBs) are intense, short pulses of gamma rays,
occurring at random positions in the sky. They emit large amounts
of energy (up to $\sim 10^{53}$~erg) and thus are detectable up to 
cosmological distances, possibly at $z = 10$ and beyond. The discovery
of high-redshift GRBs is one of the main goals of the {\it Swift} mission
(Gehrels et al. 2004) and several ground-based facilities were prepared to be able to
detect their counterparts (e.g. Chincarini et al. 2003).

GRBs are accompanied by long-lasting X-ray, optical, and radio
counterparts (afterglows).  In the first few hours after the explosion,
these are much brighter than any known quasar.  In particular, optical
and near-infrared (NIR) afterglow spectroscopy can easily provide their
redshifts and probe the intervening gas along the line of sight. The
possible biases affecting the GRB distribution are likely very different
from those affecting other classes of objects, such as AGNs, damped
Ly$\alpha$ absorbers, and galaxies. Therefore GRBs and their afterglows
provide an effective and independent way to probe the high-redshift
Universe. As a matter of fact, the observed redshift distribution of
long-duration GRBs, before the {\it Swift} era, peaks at $z \sim 1.6$
and extends up to $z = 4.5$ \citep{And00}. A significant fraction of
GRBs has been predicted to occur even at higher redshifts
\citep[e.g.][]{LR00,Goro04,Nata05,Mesi05}, during the re-ionization
epoch \citep[$6 \la z \la 20$,][]{gne97,Fan02,Sper03}. Since the
progenitors of long-duration GRBs are thought to be short-lived massive
stars \citep[]{Ga98,Sta03,Hjo03,Male04}, their detection at large
redshifts provides a direct measurement of the star formation rate in
the early Universe. Finally, high-redshift GRBs may also serve as useful
tools to study the geometry of the Universe \citep{Amati02,Ghirla04}.

Our group, the MISTICI%
\footnote{Multiwavelength Italian {\it Swift} Team with International Co-Investigators.}
collaboration, has been pursuing follow-up studies of GRB afterglows for
several years, concentrating on various aspects of the GRB phenomenon,
among them the study of their host galaxies and of foreground DLA
systems at high redshift \citep[]{Fiore05,DElia05}. Here we present
optical and NIR photometric observations of GRB\,050904. This burst was
detected by the Burst Alert Telescope \citep[BAT;][]{Bart05} onboard
{\it Swift} on 2005 September 4 at 01:51:44 UT \citep{Cumm05}. It was a
long, bright burst with an observed duration $T_{90} = 225 \pm 10$~s
\citep{Saka05}. The 15--150 keV fluence was $(5.4 \pm 0.2) \times
10^{-6}$~erg~cm$^{-2}$. The BAT spectrum can be described by a hard
power law with photon index $\approx 1.2$. Long-lasting flaring activity
was detected in the X-ray band up to 50~ks \citep{Cusu05,Watson05}.
A high redshift value for this GRB was first suggested by
\citet{Hais05a}, determined photometrically by us \citep[$z =
6.1^{+0.37}_{-0.12}$;][]{Anto05} and then confirmed spectroscopically by
\citet{Kawai05}, who measured $z = 6.29 \pm 0.01$.  A photometric
redshift has been derived also by \citet{Hais05b} and \citet{Pri05}. To
date, this is by far the most distant GRB discovered and one of the most
distant objects known in the Universe.

\begin{figure}
\includegraphics[width=\columnwidth]{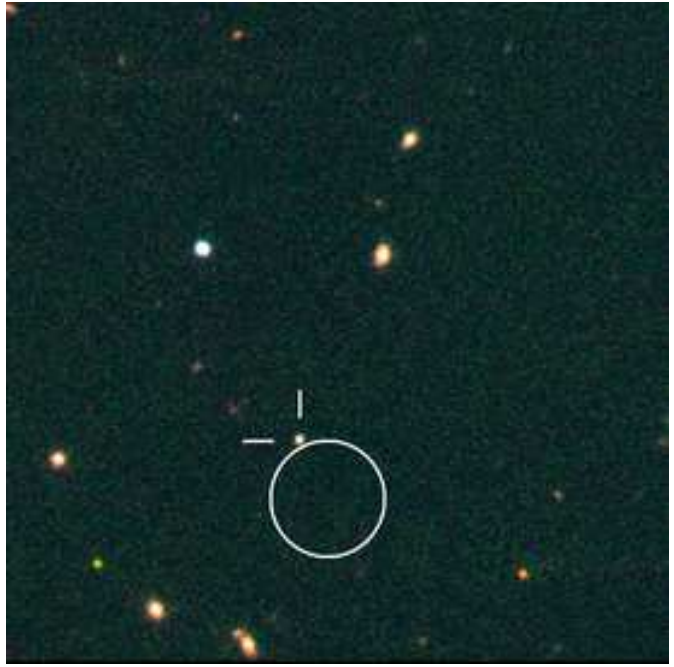}

\caption{Three colour-composite image based on our NIR bands ($J$, $H$, and $K$)
obtained with VLT+ISAAC on 5 September. The circle represents the $6\arcsec$ 
XRT afterglow error box. The lines show the position of the optical afterglow.
The field is $65\arcsec \times 65\arcsec$ wide.
North is up, East is left. 
\label{fig:ima}}
\end{figure}

\section{Observations and data analysis}\label{sec:data}

\begin{table}
\caption{Observation log and photometry of the transient source. Errors
are at the 1$\sigma$ confidence level, while upper limits are given at
3$\sigma$. Note that the CAFOS and FORS2 $I$-band filters (which we
called $I_1$ and $I_2$) are different (their central wavelengths being
8500~\AA{} and 7680~\AA, respectively). Since the Lyman $\alpha$ break
falls within this band
dropout is occuring
inside this band, this leads to a significant magnitude
difference.\label{tab:mag}}
\centering
\begin{tabular}{ccrccr}
\hline
{\bf Date}   &{\bf Instr.} &$t-t_0$ &{\bf Exp.} &{\bf Filter} &{\bf Magnitude} \\
(UT)         &             &(d)     &(min)      &             &                \\
\hline
05/09/05 04:47 &ISAAC      &1.125   &15        & $K_{\rm s}$  & 18.24$\pm$0.07 \\
05/09/06 05:23 &ISAAC      &2.146   &30        & $K_{\rm s}$  & 19.12$\pm$0.07 \\ \hline
05/09/05 04:22 &ISAAC      &1.104   &20        & $H$          & 19.09$\pm$0.07 \\
05/09/06 04:42 &ISAAC      &2.125   &30        & $H$          & 19.96$\pm$0.07 \\ \hline
05/09/05 02:31 &NICS       &1.029   &50        & $J$          & 19.58$\pm$0.14 \\
05/09/05 03:55 &ISAAC      &1.087   &20        & $J$          & 19.92$\pm$0.04 \\
05/09/06 04:01 &ISAAC      &2.092   &35        & $J$          & 20.75$\pm$0.07 \\
05/09/07 04:23 &ISAAC      &3.104   &60        & $J$          & 21.66$\pm$0.08 \\ 
05/09/08 05:44 &ISAAC      &4.162   &60        & $J$          & 21.91$\pm$0.08 \\ 
05/09/09 09:26 &ISAAC      &5.317   &60        & $J$          & 22.45$\pm$0.19 \\
05/09/11 06:18 &ISAAC      &7.183   &102       & $J$          & $>23.2$        \\ \hline
05/09/05 02:25 &WFC        &1.021   &28        & $z$          & 21.50$\pm$0.25 \\ 
05/09/05 07:15 &FORS2      &1.250   &90        & $z$          & 21.80$\pm$0.17 \\ \hline
05/09/04 23:17 &CAFOS      &0.892   &50        & $I_1$        & 21.89$\pm$0.20 \\
05/09/05 01:35 &CAFOS      &0.987   &50        & $I_1$        & 22.03$\pm$0.20 \\
05/09/05 03:38 &CAFOS      &1.075   &50        & $I_1$        & 22.40$\pm$0.20 \\
05/09/05 06:54 &FORS2      &1.208   &60        & $I_2$        & 24.10$\pm$0.18 \\ \hline
05/09/05 01:30 &CAFOS      &0.983   &90        & $R$          & $>24.1$        \\ \hline
05/09/05 02:48 &LAICA      &1.038   &30        & $V$          & $>24.1$        \\ \hline \hline
\end{tabular}
\end{table}

\begin{figure}
\includegraphics[width=\columnwidth]{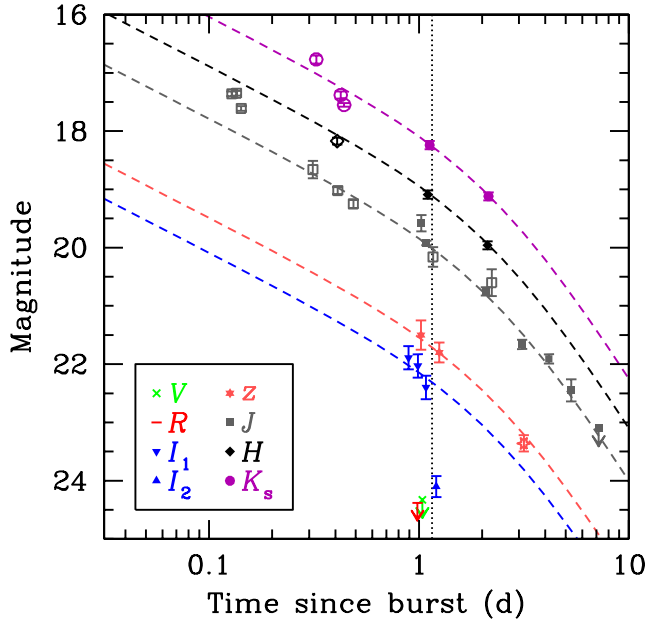} 
\caption{Light curve of the afterglow of GRB\,050904, together with the
best fit (dashed lines) computed in the $J$ band. The same fit is
reported for the other bands. Open symbols represent data from Haislip
et al. (2005b).  The vertical dotted line marks the epoch at which we
computed the spectral energy distribution (Fig. 3).\label{fig:lc}}
\end{figure}

Our observations of the field of GRB\,050904 were mostly performed with
the 8.2m ESO Very Large Telescope (VLT), equipped with the ISAAC and
FORS2 instruments (Table~\ref{tab:mag}). Further data were collected
with the 2.2m and 3.5m Calar Alto telescopes (equipped with CAFOS and
LAICA, respectively), the 2.5m Isaac Newton Telescope (equipped with
WFC), and the 3.6m Telescopio Nazionale Galileo (equipped with NICS). A
NIR counterpart to GRB\,050904 was first reported by Haislip et
al. (2005a).
In our first images, taken on 2005 Sep 4/5, this object was not
present. A pointlike source was however clearly visible in the
$IzJHK_{\rm s}$ bands, at the edge of the XRT error box
\citep{Mineo05}. Its coordinates were $\alpha_{\rm J2000} = 00^\mathrm{h}
54^\mathrm{m} 50\fs83$, $\delta_{\rm J2000} = +14\degr 05\arcmin
10\farcs0$. We suggested this source to be the afterglow of GRB\,050904
(D'Avanzo et al., 2005; see also Fig.~\ref{fig:ima}). \citet{Nyse05}
subsequently provided revised coordinates for the object of \citet{Hais05a},
showing it was coincident with our candidate.
The afterglow was also detected by the TAROT robotic telescope up to
eight minutes after the GRB (Klotz et al. 2005).

Data reduction was carried out following standard procedures. Aperture
photometry was performed using SExtractor (Bertin \& Arnouts 1996).
Flux calibration was achieved by observing Landolt standard fields ($I$
band) and against the SDSS ($z$ band) and 2MASS (NIR bands) surveys.

In Fig.~\ref{fig:lc} we show our data, together with those by Haislip et
al. (2005b). The light curve in the $J$ band, for which we have the best
coverage, shows a complex behaviour. Haislip et al. (2005b) have
reported a flattening at early times, suggesting that the early emission
could be related to a different component, possibly the reverse
shock. Our data extend for significantly longer and show a further
steepening. Excluding the data at $t < 0.4$~d, we can model the whole
dataset with a smoothly broken power law: $F(t) = 2F_{\rm b} /
[(t/t_{\rm b})^{\alpha_1} + (t/t_{\rm b})^{\alpha_2}]$, where
$\alpha_1$, $\alpha_2$ are the early- and late-time slopes, $t_{\rm b}$
is the break time and $F_{\rm b}$ is the flux at $t_{\rm b}$.  
The fit provides $\alpha_1 = 0.72^{+0.15}_{-0.20}$, $\alpha_2 = 2.4 \pm 0.4$,
$t_{\rm b} = 2.6 \pm 1.0$~d. Errors are at the 1$\sigma$ confidence
level throughout the paper.

\begin{figure}
\includegraphics[width=\columnwidth]{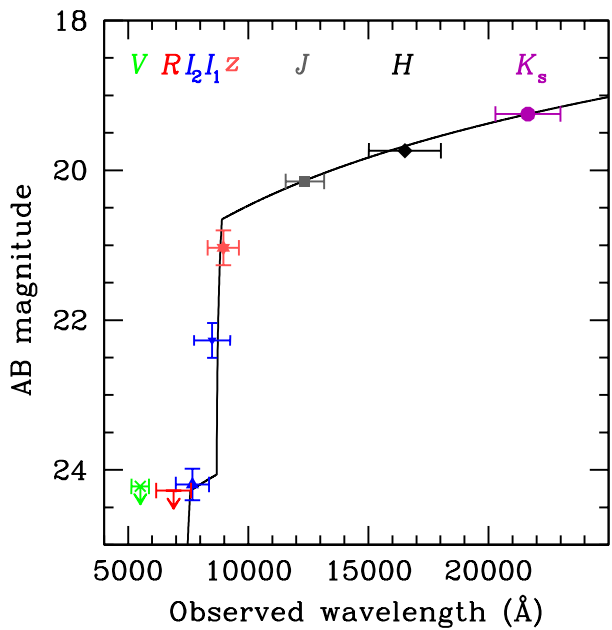}
\caption{Photometric spectral energy distribution of the
afterglow. Magnitudes were reported at a common epoch (1.155~d after the
GRB) using the measured decay law. The solid line shows the best fit
model for the afterglow. Data are corrected for Galactic extinction
($A_V = 0.21$~mag).\label{fig:SED}}
\end{figure}

Figure~\ref{fig:SED} shows the spectral energy distribution of the
afterglow 1.155~d after the GRB (the epoch around which our measurements
cluster). A marked dearth of flux is observed blueward of the $J$
passband, while the $J-H$ and $H-K$ colors are not particularly
red. Dust extinction cannot produce such a sharp cutoff, so that a
redshifted Lyman dropout must be invoked (due to the strong absorption
by neutral Hydrogen at wavelengths shorter than Ly$\alpha$). The
redshift of GRB\,050904 was computed using two different programs.  The
first is a code based on a widely used technique to derive photometric
redshifts of galaxies and quasars (e.g. Fontana et al. 2000), modified
to take into account the afterglow properties. In this case we computed a
library of synthetic GRB afterglow spectra at arbitrary redshifts,
modeled with power laws ($F_\nu \propto\nu^{-\beta}$). Reporting all our
photometric measurements at a common epoch (adopting the observed decay
law), the best fit gives $z = 6.30 \pm 0.07$ and $\beta=1.25 \pm 0.25$.
The second is the {\it z-ph-REM} code \citep{FS04}, specifically
designed for GRB afterglows and based on the work by Fern\'andez-Soto et
al. (1999).  This code assumes that the temporal and spectral dependence
of the afterglow flux follows a power law ($F_\nu \propto t^{-\alpha}
\nu^{-\beta}$). In this case, also including the UVOT optical limits
\citep{Cucch05}, the best fit provided $z = 6.3 \pm 0.1$ and $\beta =
1.2 \pm 0.3$. Both codes use a $\chi ^2$--minimisation technique to find
the best-fitting spectral template to the observed colors.
We note that the spectral index is significantly redder than that reported 
by \citet{Pri05}, which was however measured at an earlier epoch.

\section{Discussion}\label{sec:disc}

Despite its high redshift, the optical afterglow of GRB\,050904 is not
peculiar with respect to other GRBs. For example, its $R$-band spectral
luminosity extrapolated at $t = 12$~hr (rest-frame time) is%
\footnote{Assuming a cosmology with $\Omega_{\rm m} = 0.27$,
$\Omega_\Lambda = 0.73$, and $h_0 = 0.71$.}
$L_\nu \sim 10^{31}$~erg~s$^{-1}$~Hz$^{-1}$, consistent with the values
typically observed for previous GRBs \citep{Nardini05,li05}. Its
spectral and temporal indices are also typical among GRB afterglows. In
particular, the afterglow colours leave small room for dust extinction,
if any (note that we are observing the rest-frame ultraviolet band).
The precise gamma-ray energy budget of this burst is not known, since
the observed peak spectral energy  lies outside the
sensitivity BAT range. Cusumano et al. (2005) infer that the
isotropic-equivalent gamma-ray budget $E_{\gamma,\rm iso}$ was between
$6.6 \times 10^{53}$ and $3.2 \times 10^{54}$~erg, depending on the
bolometric correction ($E_{\gamma,\rm iso} \simeq 6 \times 10^{52}$~erg
in the 20-100 keV rest-frame band).
The steepening in the light curve may be due to a jet break. In this
case, adopting the standard formalism \citep{SPH99}, we can infer a jet
half-opening angle $\vartheta_{\rm jet} \sim 3\degr$,
assuming a radiative efficiency $\eta = 20\%$ and a circumburst medium
density $n = 3$~cm$^{-3}$ (as in Ghirlanda et al. 2004).  In this case
the beaming-corrected energy would be $E_\gamma \sim (4 \div 12) \times
10^{51}$~erg.
For this burst the rest-frame peak energy is constrained to be $E_{\rm
p} \ga 1100$~keV. In the planes $E_{\rm p}$ vs $E_{\gamma,\rm iso}$ and
$E_{\rm p}$ vs $E_\gamma$, the derived limits show that GRB\,050904 is
consistent with both the Amati and Ghirlanda relations
\citep{Amati02,Ghirla04}.

The optical/NIR afterglows of GRBs detected by {\it Swift} are dimmer
than those of other missions, on the average by 2.5 mag at $t = 12$~hr
(Berger et al. 2005). More than one third have no optical counterpart,
despite having a well-localized X-ray afterglow. For a good fraction of
these bursts, this could be due either to a high redshift and/or to dust
extinction (e.g. Zhang \& M{\'e}sz{\'a}ros 2004 and references therein;
see also Filliatre et al. 2005).  In about 10 months of operations, the
number of {\it Swift} long-duration GRBs with measured redshift has been
measured is 16, while the number of non-{\it Swift} GRBs with redshift
is 43 (obtained over a time period of 7 years). The redshift
distribution of the pre-{\it Swift} bursts has a peak at $z
\sim 1.6$, extending to higher values (the record was $z = 4.50$;
Andersen et al. 2000). Although the number is still small, the 16 {\it
Swift} long-duration GRBs with measured redshift have a statistically
higher redshift, with an average value of 2.8 (with one object at $z > 6$,
two with $4 < z < 6$, three with $3 < z < 4$ and four with $2 < z < 3$,
e.g. Jakobsson et al. 2005).
It is very likely that some of the {\it Swift} bursts without
optical counterpart are also at high redshift.  Moreover, the number of
bursts for which it is possible to perform low- and high-resolution
spectroscopic studies at moderate redshift ($z > 2$) is rapidly
increasing (e.g. D'Elia et al. 2005; Chen et al. 2005).

Finally we note that a GRB survey is much more efficent in finding
high-redshift objects than those targetting other sources.  For instance
only seven quasars out of $\sim 76\,000$ (selected from a sample of
$\sim 850\,000$ sources for which spectra were acquired, over a 4783
deg$^2$ region of the sky) were detected in the Sloan Digital Sky Survey%
\footnote{\texttt{http://www.sdss.org/dr4/}\,.}
at $z > 6$.  For comparison, {\it Swift} is monitoring about 1.4~sr of
the sky at any time, and it has been able to catch one event at $z > 6$,
out of a sample of 16 with measured redshift, selected from a sample of
$\sim 60$ GRBs.

\section{Conclusions}

We observed the afterglow of GRB\,050904 in the $VRIzJHK_{\rm s}$
filters. We derived a photometric redshift $z = 6.3 \pm 0.1$ using two
independent techniques. This redshift has been estimated by the Lyman
break falling between the $I$ and $J$ filters, therefore it is very
robust and has been confirmed spectroscopically by
\citet{Kawai05}, who measured $z = 6.29 \pm 0.01$.

The high redshift of GRB\,050904 opens the question of whether its
progenitor was a Pop III star (see, for example, Woosley \& Heger 2004;
Scannapieco et al. 2005). In this case, however, it would be difficult
for the star to lose its envelope. Wind-driven mass loss depends upon
metallicity, and it scales as $\dot{M} \propto Z^{0.5\div 0.9}$
(e.g. Nugis \& Lamers 2000; Kudritzki 2002; Vink \& de Koter 2005). 
The discovery of GRB\,050904, possibly in a low metallicity environment 
like the one characterizing the early Universe, hints at the need for
a close companion that can efficiently strip off the envelope of the
progenitor. However, the efficiency of producing GRBs by Pop III stars
is currently unknown and may be very low (e.g. Bromm \& Loeb 2005).

A more convincing alternative is the possibility that the progenitor of
GRB\,050904 was a massive, non-pristine star. This implies that star
formation was already active at $z \ga 6$, and that metal enrichment had
already started. This finding is consistent with recent measurements of
the comoving luminosity density of star-forming galaxies
\citep{gia04}, which show that the star formation rate varies only
slowly with redshift over the range $2 < z < 6$ \citep[see
also][]{lefev05}. These results are nevertheless still subject to
important -- and not well known -- corrections due to incompleteness,
dust absorption, and systematic effects. GRBs are not affected by dust
absorption, and are thus in principle very effective in tracing star
formation even in this high redshift range.

{\it Swift} has been able to catch one GRB at $z \sim 6$, out of a
sample of 16 events for which the redshifts have been measured.  By
using the whole sample of about 60 GRBs discovered by {\it Swift}, we
find, as a simple application of Poisson statistics in the small-number
regime (Gehrels 1986), that $4^{+9}_{-3}$ events are expected to be
discovered at $z>6$. A fraction of about $4/60 \sim 7\%$, as we have
derived on empirical grounds, is consistent with the theoretical
estimate recently provided by Bromm \& Loeb (2005), who predicted that
about 10\% of the events discovered by {\it Swift} are at $z>6$.  Once
we will be able to measure a significant number of high GRB redshifts,
we will be in a position to sample the star formation rate of the
Universe within the first Gyr from the Big Bang.

\begin{acknowledgements}
This research was supported at OAB and OAR by ASI grant I/R/039/04.
We thank A. Fontana for allowing us to use his photometric redshift code.
We acknowledge the excellent support of the ESO, TNG and CAHA staff. We also 
thank the anonymous referee for her/his prompt reply and for useful comments.
\end{acknowledgements}

\end{document}